\documentclass[times,twocolumn,final]{elsarticle}
\usepackage{prletters}

\usepackage[utf8]{inputenc}

\usepackage{amsthm}
\usepackage[dvipsnames]{xcolor}
\usepackage{amsfonts} 					
\usepackage{amsmath}			
\usepackage{stmaryrd}					
\usepackage{mathtools} 
\usepackage{nicefrac} 
\usepackage{amssymb}					
\usepackage{booktabs}					
\usepackage[vlined,ruled,linesnumbered]{algorithm2e}
\SetKwFor{Forever}{forever}{do}{endforever}
\usepackage{etoolbox}
\usepackage{graphicx}					
\usepackage{longtable,rotating}			
\usepackage{siunitx}			
\robustify\bfseries
\usepackage{wasysym}					
\usepackage{url} 
\usepackage{microtype} 
\usepackage{xspace} 
\usepackage{csquotes} 
\usepackage[colorlinks=true,linkcolor=blue]{hyperref}
\usepackage{cleveref}
\usepackage{nameref}
\allowdisplaybreaks
\crefname{ineq}{inequality}{inequalities}
\Crefname{ineq}{Inequality}{Inequalities}
\creflabelformat{ineq}{#2{(#1)}#3}
\crefname{equation}{equation}{equations}
\usepackage{IEEEtrantools}
\makeatletter
\patchcmd{\@IEEEyesnumber}
  {\stepcounter}
  {\refstepcounter}
  {}{}
\patchcmd{\@@IEEEeqnarray}
  {\stepcounter}
  {\refstepcounter}
  {}{}
\patchcmd{\@@IEEEeqnarraycr}
  {\stepcounter{IEEEsubequation}}
  {\refstepcounter{IEEEsubequation}}
  {}{}
\patchcmd{\@@IEEEeqnarraycr}
  {\stepcounter{IEEEsubequation}}
  {\refstepcounter{IEEEsubequation}}
  {}{}
\patchcmd{\@@IEEEeqnarraycr}
  {\stepcounter{IEEEequation}}
  {\refstepcounter{IEEEequation}}
  {}{}
\patchcmd{\@@IEEEeqnarraycr}
  {\stepcounter{IEEEequation}}
  {\refstepcounter{IEEEequation}}
  {}{}
\makeatother

\usepackage{tikz,pgf,pgfplots,pgfplotstable}
\usetikzlibrary{calc,backgrounds,arrows,decorations.markings,decorations.pathreplacing,matrix,shapes,graphs,positioning,fit,snakes}
\usepgfplotslibrary{groupplots}
\pgfplotsset{compat=1.6}
\usepackage{ifthen}
\usepackage{xstring}
\usepackage{calc}
\usepackage{pgfopts}
\makeatletter
\newcommand{\gettikzxy}[3]{%
  \tikz@scan@one@point\pgfutil@firstofone#1\relax
  \edef#2{\the\pgf@x}%
  \edef#3{\the\pgf@y}%
}
\makeatother
\usepackage{forest}

\newcommand{\ie}{i.\,e.\@\xspace}

\newcommand{\wrt}{w.\,r.\,t.\@\xspace}

\newcommand{\cf}{cf.\@\xspace}

\newcommand{\NP}{\ensuremath{\mathcal{NP}}\xspace}

\newcommand{\REFINE}{\texttt{REFINE}\xspace}

\newcommand{\ALG}{\texttt{ALG}\xspace}
\newcommand{\KREFINE}[1][K]{\texttt{#1-REFINE}\xspace}

\newcommand{\IPFP}{\texttt{IPFP}\xspace}

\newcommand{\BPBEAM}{\texttt{BP-BEAM}\xspace}
\newcommand{\IBPBEAM}{\texttt{IBP-BEAM}\xspace}
\newcommand{\RANDPOST}{\texttt{RANDPOST}\xspace}

\newcommand{\SWAPCOST}{\ensuremath{\operatorname{\mathtt{SWAP-COST}}}\xspace}
\newcommand{\SWAP}{\ensuremath{\operatorname{\mathtt{SWAP}}}\xspace}
\newcommand{\GENERATENODEMAPS}{\ensuremath{\operatorname{\mathtt{GEN-NODE-MAPS}}}\xspace}
\newcommand{\UPDATESCORES}{\ensuremath{\operatorname{\mathtt{UPD-SCORES}}}\xspace}

\newcommand{\mutagenicity}{\textsc{muta}\xspace}
\newcommand{\mutagenicityl}{\textsc{muta-l}\xspace}
\newcommand{\mutagenicityn}{\textsc{muta-n}\xspace}
\newcommand{\protein}{\textsc{protein}\xspace}
\newcommand{\grec}{\textsc{grec}\xspace}
\newcommand{\fingerprint}{\textsc{fp}\xspace}

\newcommand{\LSAPE}{\relax\ifmmode\operatorname{LSAPE}\else{LSAPE}\fi\xspace}
\newcommand{\LSAP}{\relax\ifmmode\operatorname{LSAP}\else{LSAP}\fi\xspace}
\newcommand{\QAP}{\relax\ifmmode\operatorname{QAP}\else{QAP}\fi\xspace}
\newcommand{\QAPE}{\relax\ifmmode\operatorname{QAPE}\else{QAPE}\fi\xspace}
\newcommand{\IP}{\relax\ifmmode\mathrm{IP}\else{IP}\fi\xspace}
\newcommand{\LP}{\relax\ifmmode\mathrm{LP}\else{LP}\fi\xspace}
\newcommand{\MIP}{\relax\ifmmode\mathrm{MIP}\else{MIP}\fi\xspace}
\newcommand{\GED}{\relax\ifmmode\operatorname{GED}\else{GED}\fi\xspace}
\newcommand{\BED}{\relax\ifmmode\operatorname{BED}\else{BED}\fi\xspace}
\newcommand{\FBED}{\relax\ifmmode\operatorname{FBED}\else{FBED}\fi\xspace}
\newcommand{\UB}{\ensuremath{\mathit{UB}}\xspace}
\newcommand{\LB}{\ensuremath{\mathit{LB}}\xspace}
\newcommand{\GT}{\relax\ifmmode\mathrm{GT}\else{GT}\fi\xspace}
\newcommand{\SGI}{\relax\ifmmode\mathrm{SGI}\else{SGI}\fi\xspace}
\newcommand{\GI}{\relax\ifmmode\mathrm{GI}\else{GI}\fi\xspace}

\newcommand{\R}{\ensuremath{\mathbb{R}}\xspace}

\newcommand{\N}{\ensuremath{\mathbb{N}}\xspace}
\newcommand{\C}{\ensuremath{\mathbf{C}}\xspace}
\newcommand{\D}{\ensuremath{\mathbf{D}}\xspace}

\newcommand{\B}{\ensuremath{\mathbf{B}}\xspace}

\newcommand{\X}{\ensuremath{\mathbf{X}}\xspace}
\newcommand{\M}{\ensuremath{\mathbf{M}}\xspace}

\newcommand{\tp}[1]{#1^\mathsf{T}}

\DeclareMathOperator*{\argmin}{arg\,min}

\DeclareMathOperator*{\vect}{vec}

\DeclareMathOperator*{\maxdeg}{max\,deg}
\DeclarePairedDelimiter\ceil{\lceil}{\rceil}

\newcommand{\x}{\ensuremath{\mathbf{x}}\xspace}

\newdefinition{definition}{Definition}
\newtheorem{proposition}{Proposition}

\tikzstyle{2REFINE}=[mark=square, mark size=2pt, color=NavyBlue]
\tikzstyle{3REFINE}=[mark=triangle,mark size=2pt, color=Purple]
\tikzstyle{IPFP}=[mark=asterisk,mark size=2pt, color=Green]
\tikzstyle{BPBEAM}=[mark=x,mark size=2pt, color=Orange]
\tikzstyle{IBPBEAM}=[mark=+, mark size=2pt, color=Red]

\newcommand{\addplots}[5]{
\nextgroupplot[
ymode={#5},
title={#1}, 
legend to name=grouplegend, 
ylabel={#4},
xlabel={$L$},
title style={yshift=-1ex,},
xlabel style={yshift=1ex,font=\footnotesize},
ylabel style={yshift=-1ex,font=\footnotesize},
xticklabel style = {yshift=.5ex,font=\tiny},
yticklabel style = {xshift=.5ex,font=\tiny},
]
\addlegendimage{2REFINE, only marks};
\addlegendimage{3REFINE, only marks};
\addlegendimage{IPFP, only marks};
\addlegendimage{BPBEAM, only marks};
\addlegendimage{IBPBEAM, only marks};
\addplot[2REFINE] table [x=L, y=2REFINE_#3, col sep=comma, ignore chars={"}] {#2};
\addplot[3REFINE] table [x=L, y=3REFINE_#3, col sep=comma, ignore chars={"}] {#2};
\addplot[IPFP] table [x=L, y=IPFP_#3, col sep=comma, ignore chars={"}] {#2};
\addplot[BPBEAM] table [x=L, y=BPBEAM_#3, col sep=comma, ignore chars={"}] {#2};
\addplot[IBPBEAM] table [x=L, y=IBPBEAM_#3, col sep=comma, ignore chars={"}] {#2};
\legend{\KREFINE[2],\KREFINE[3],\IPFP,\BPBEAM,\IBPBEAM};
}

\journal{Pattern Recognition Letters}

\begin{document}

\begin{frontmatter}
\title{Improved local search for graph edit distance}

\author[1]{Nicolas Boria}
\author[2]{David B. Blumenthal}
\author[1]{S\'{e}bastien Bougleux}
\author[1]{Luc Brun}

\address[1]{Normandie Univ, ENSICAEN, UNICAEN, CNRS, GREYC, 14000 Caen, France}
\address[2]{Free University of Bozen-Bolzano, Faculty of Computer Science, Piazza Dominicani 3, 39100 Bolzano, Italy}

\begin{abstract}
The graph edit distance (\GED) measures the dissimilarity between two graphs as the minimal cost of a sequence of elementary operations transforming one graph into another. This measure is fundamental in many areas such as structural pattern recognition or classification. However, exactly computing \GED is \NP-hard. Among different classes of heuristic algorithms that were proposed to compute approximate solutions, local search based algorithms provide the tightest upper bounds for \GED. In this paper, we present \KREFINE and \RANDPOST. \KREFINE generalizes and improves an existing local search algorithm and performs particularly well on small graphs. \RANDPOST is a general warm start framework that stochastically generates promising initial solutions to be used by any local search based \GED algorithm. It is particularly efficient on large graphs. An extensive empirical evaluation demonstrates that both \KREFINE and \RANDPOST perform excellently in practice.
\end{abstract}


\end{frontmatter}

\section{Introduction}\label{sec:intro}

In many areas such as pattern recognition or graph classification, computing a graph (dis-)similarity measure is a central issue \cite{conte:2004aa,foggia:2014aa,vento:2015aa,brun:2018aa}. One popular approach is to embed the graphs into multi-dimensional vector spaces and then to compare their vector representations \cite{bai:2009aa,xiao:2009aa,bai:2010aa}. This approach has the advantage that is allows fast computations. The main drawback is that embedding the graphs into vector spaces always comes at the price of information loss. 

If local information that would be lost by the embeddings is crucial, it is recommendable to compare the graphs directly in the graph space. The graph edit distance (\GED) is one of the most widely used measures for this purpose \cite{bunke:1983aa,sanfeliu:1983aa}. \GED is defined as the minimum cost of an edit path transforming one graph into another, where an edit path is a sequence of node and edge insertions, deletions, and substitutions. Equivalently, it can be defined as the minimum cost of an edit path induced by a node map that assigns nodes of the source graph to nodes of the target graph \cite{blumenthal:2018ab}.

As exactly computing \GED is \NP-hard \cite{zeng:2009aa}, research has mainly focused on heuristics \cite{blumenthal:2019ab}. The development of heuristics was particularly triggered by the algorithms presented in \cite{riesen:2009aa} and \cite{zeng:2009aa}, which use transformations to the linear sum assignment problem with error correction (\LSAPE) \cite{bougleux:2018aa}\,---\,a variant of the linear sum assignment problem (\LSAP) where rows and columns may also be inserted and deleted\,---\,to compute upper bounds for \GED. Further transformations from \GED to \LSAPE have been proposed in \cite{zheng:2015aa,blumenthal:2018aa,blumenthal:2018ad,carletti:2015aa,gauzere:2014aa,blumenthal:2017aa,riesen:2014ad,cortes:2015aa}.

\LSAPE based heuristics are typically quite fast but yield loose upper bounds on some graphs. Tighter upper bounds for \GED can be obtained by algorithms that use variants of local search. Given one or several initial node maps, local search based algorithms explore suitably defined neighborhoods to find improved node maps that induce cheaper edit paths. Several algorithms have been proposed that instantiate this paradigm: \REFINE \cite{zeng:2009aa} varies the initial node maps via binary swaps; \IPFP \cite{bougleux:2016aa,bougleux:2017aa,blumenthal:2018ac} computes locally optimal node maps by using a variant of the classical Frank-Wolfe algorithm \cite{frank:1956aa,leordeanu:2009aa}; \BPBEAM \cite{riesen:2014aa} uses beam search to improve the initial node maps; and \IBPBEAM \cite{ferrer:2015ab} further improves \BPBEAM by iteratively running it with different processing orders. 

In this paper, we propose a new local search based algorithm \KREFINE and a warm start framework \RANDPOST. \KREFINE generalizes and improves the existing algorithm \REFINE in three respects: Firstly, \KREFINE considers not only binary swaps, but rather swaps of size up to $K$, where $K\in\N_{\geq2}$ is a meta-parameter. Secondly, \KREFINE computes the swap costs more efficiently than \REFINE, which leads to a significant gain in runtime performance. Thirdly, unlike \REFINE, \KREFINE allows the improved node map to contain fewer node substitutions than the original node map, which tightens the produced upper bound. 

The framework \RANDPOST extends the local search paradigm and hence improves all local search based \GED algorithms. In a first step, \RANDPOST runs a local search algorithm from a set of initial node maps. Subsequently, \RANDPOST stochastically generates a new set of initial node maps from the converged solutions, and then re-runs the local search algorithm. This process iterates until a user-specified number of iterations has been reached. Extensive experiments show that \KREFINE performs extremely well on small to medium sized graphs, while using \RANDPOST is particularly effective on larger graphs.

The remainder of this paper is organized as follows: In \Cref{sec:prelim}, we fix concepts and notations and present the general local search framework. In \Cref{sec:related}, we summarize existing local search algorithms. In \Cref{sec:k-refine} and \Cref{sec:randpost}, we present \KREFINE and \RANDPOST. In \Cref{sec:experiments}, \KREFINE and \RANDPOST are evaluated empirically. \Cref{sec:conclusions} concludes the paper. The paper extends the results published in \cite{boria:2018aa}, where a preliminary version of \RANDPOST was presented.

\section{Preliminaries}\label{sec:prelim}
\subsection{Definitions and notations}\label{sec:prelim:notations}

An \emph{undirected labeled graph} $G$ is a $4$-tuple $G=(V^G,E^G,\allowbreak\ell^G_V,\ell^G_E)$, where $V^G$ and $E^G$ are sets of nodes and edges, $\Sigma_V$ and $\Sigma_E$ are label alphabets, and $\ell^G_V:V^G\to\Sigma_V$, $\ell^G_E:E^G\to\Sigma_E$ are labeling functions. The \emph{dummy symbol} $\epsilon$ denotes dummy nodes and edges as well as their labels. Throughout the paper, we denote the nodes of a graph $G$ by $V^G\coloneqq\{u_i\mid i\in[|V^G|]\}$ and the nodes of a graph $H$ by $V^H\coloneqq\{v_k\mid k\in[|V^H|]\}$, where the index set $[N]$ is defined as $[N]\coloneqq\{n\in\N\mid 1\leq n\leq N\}$, for all $N\in\N$.


We denote by \emph{assignment} a pair $(u,v)$ in $(V^G \cup \{\epsilon\})\times(V^H \cup \{\epsilon\})$, and we define a \emph{node map} $\pi$ as a set of assignments such that each node in $V^G$ and $V^H$ appears in exactly one assignment of $\pi$. The notation $(u,v)\in\pi$ is considered equivalent to both $\pi(u)=v$ and $\pi^{-1}(v)=u$, and $\Pi(G,H)$ denotes the set of all node maps between $G$ and $H$. For edges $e=(u,u^\prime)\in E^G$ and $f=(v,v^\prime)\in E^H$, we introduce the short-hand notations $\pi(e)\coloneqq(\pi(u),\pi(u^\prime))$ and $\pi^{-1}(f)\coloneqq(\pi^{-1}(v),\pi^{-1}(v^\prime))$.

A node map $\pi\in\Pi(G,H)$ specifies for all nodes and edges of $G$ and $H$ whether they are substituted, deleted, or inserted. Each of these operations has an induced cost. An assignment $(u,v)$ induces a node substitution with cost $c_V(u,v)$ if $u\neq\epsilon$ and $v\neq\epsilon$, a node deletion with cost $c_V(u,\epsilon)$ if $u\neq\epsilon$ and $v=\epsilon$, and a node insertion with cost $c_V(\epsilon,v)$ if $u=\epsilon$ and $v\neq\epsilon$. Similarly, a pair of assignments $((u,v),(u',v'))$ induces an edge substitution with cost $c_E((u,u^\prime),(v,v'))$ if $(u,u^\prime)\in E^G$ and $(v,v')\in E^H$, an edge deletion with cost $c_E((u,u^\prime),\epsilon)$ if $(u,u^\prime)\in E^G$ and $(v,v')\notin E^H$, and an edge insertion with cost $c_E(\epsilon,(v,v'))$ if $(u,u^\prime)\notin E^G$ and $(v,v')\in E^H$. The edit cost functions $c_V$ and $c_E$ are defined in terms of the labeling functions $\ell^G_V$, $\ell^H_V$, $\ell^G_E$, and $\ell^H_E$, \ie, nodes and edges with the same labels induce the same edit costs.

Any node map $\pi \in \Pi(G,H)$ hence induces an \emph{edit path} $P_\pi$ between $G$ and $H$. The cost $c(P_\pi)$ of $P_\pi$ is given as follows:

\begin{IEEEeqnarray*}{rCl}
c(P_\pi)&=&\underbrace{\sum_{\substack{u\in V^G\\\pi(u)\in V^H}}c_V(u,\pi(u))}_{\text{node substitutions}}+\underbrace{\sum_{\substack{u\in V^G\\\pi(u)\notin V^H}}c_V(u,\epsilon)}_{\text{node deletions}}+\underbrace{\sum_{\substack{v\in V^H\\\pi^{-1}(v)\notin V^G}}c_V(\epsilon,v)}_{\text{node insertions}}\\
&&+\>\underbrace{\sum_{\substack{e\in E^G\\\pi(e)\in E^H}}c_V(e,\pi(e))}_{\text{edge substitutions}}+\underbrace{\sum_{\substack{e\in E^G\\\pi(e)\notin E^H}}c_E(e,\epsilon)}_{\text{edge deletions}}+\underbrace{\sum_{\substack{f\in E^H\\\pi^{-1}(f)\notin E^G}}c_E(\epsilon,f)}_{\text{edge insertions}}
\end{IEEEeqnarray*}
We can now formally define \GED.
\begin{definition}[\GED]\label{def:ged}
The \emph{graph edit distance} (\GED) between two graphs $G$ and $H$ is defined as $\GED(G,H)\coloneqq\min_{\pi\in\Pi(G,H)}c(P_\pi)$.
\end{definition}


\subsection{Upper bounding \GED via local search}\label{sec:prelim:paradigm}

By \Cref{def:ged}, each node map $\pi\in\Pi(G,H)$ induces an upper bound $\UB\coloneqq c(P_\pi)$ for $\GED(G,H)$. Hence, a straightforward application of the local search paradigm to the problem of upper bounding \GED works as follows: Given an initial node map $\pi\in\Pi(G,H)$, run a local search algorithm to obtain an improved node map $\pi^\prime$ with $c(P_{\pi^\prime})\leq c(P_\pi)$. 

With this approach, the quality of the obtained node map $\pi^\prime$ clearly depends a lot on the initial node map $\pi$. In order to reduce this dependency, it was suggested in \cite{daller:2018aa} to generate $\kappa$ different initial solutions, run the local search algorithm on each of them (possibly in parallel), and return the best among the $\kappa$ computed local optima. In order to reduce the computing time when parallelization is available, it was suggested in \cite{boria:2018aa} to run in parallel more local searches than the number of desired local optima and to stop the whole process once the number local searches that have converged has reached the number of desired local optima. In this context, the framework runs with two parameters: $\kappa$ represents the number of initial solutions, and $0<\rho\leq1$ is defined such that $\ceil*{\rho\cdot \kappa}$ represents the number of desired computed local optima.

\section{Existing local search algorithms for \GED}\label{sec:related}

\subsection{The algorithm \REFINE}\label{sec:related:refine}

Given an initial node map $\pi\in\Pi(G,H)$, the algorithm \REFINE \cite{zeng:2009aa} proceeds as follows: Let $((u_s,v_s))^{|\pi|}_{s=1}$ be an arbitrary ordering of the initial node map $\pi$, and $G_\pi\coloneqq(V^G_\pi\cup V^H_\pi,A_\pi)$ be an auxiliary directed bipartite graph, where $V^G_\pi\coloneqq\{u_s\mid s\in[|\pi|]\}$, $V^H_\pi\coloneqq\{v_s\mid s\in[|\pi|]\}$, and $A_\pi\coloneqq\pi\cup\{(v_s,u_{s^\prime})\mid (s,s^\prime)\in[|\pi|]\times[|\pi|]\land s\neq s^\prime\}$. In other words, $G_\pi$ contains a forward arc for each assignment contained in $\pi$, and backward arcs between nodes in $V^G_\pi$ and $V^H_\pi$ that are not assigned to each other by $\pi$. A directed cycle $C\subseteq A_\pi$ in $G_\pi$ with $|C|=4$ is called swap.

For each swap $C=\{(u_s,v_s),(v_s,u_{s^\prime}),(u_{s^\prime},v_{s^\prime}),(v_{s^\prime},u_s)\}$, \REFINE checks if the swapped node map $\pi^\prime\coloneqq(\pi\setminus\{(u_s,v_s),(u_{s^\prime},v_{s^\prime})\})\cup\{(u_s,v_{s^\prime}),(u_{s^\prime},v_s)\}$ induces a smaller upper bound than $\pi$. If, at the end of the for-loop, a node map has been found that improves the upper bound, $\pi$ is updated to the node map that yields the largest improvement and the process iterates. Otherwise, the current node map $\pi$ is returned.

\subsection{The algorithm \IPFP}\label{sec:related:ipfp}

The algorithm \IPFP \cite{leordeanu:2009aa} is a variant of the Frank-Wolfe algorithm \cite{frank:1956aa} for cases where an integer solution is required. Its adaptation to \GED suggested in \cite{bougleux:2016aa,bougleux:2017aa,blumenthal:2018ac} implicitly constructs a matrix $\D\in\R^{((|V^G|+1)\cdot(|V^H|+1))\times((|V^G|+1)\cdot(|V^H|+1))}$ such that $\min_{\X\in\Pi^\prime(G,H)}\tp{\vect(\X)}\D\vect(\X)=\GED(G,H)$, where $\Pi^\prime(G,H)\subseteq\{0,1\}^{(|V^G|+1)\times(|V^H|+1)}$ contains all node maps between $G$ and $H$ in matrix form and $\vect(\cdot)$ is a vectorization operator. 

Let the cost function $c:[0,1]^{(|V^G|+1)\times(|V^H|+1)}\to \R$ be defined as $c(\X)\coloneqq\tp{\vect(\X)}\D\vect(\X)$. Starting from an initial node map $\X_0\in\Pi^\prime(G,H)$ with induced upper bound $\UB\coloneqq c(\X_0)$, the algorithm initializes $\X^\prime\coloneqq\X_0$ and converges to a, possibly fractional, local minimum by repeating the five following steps:

\begin{enumerate}
\item Populate \LSAPE instance $\C_k\coloneqq\D\vect(\X_k)$.\label{ipfp-1}
\item Compute $\B_{k+1}\in\argmin_{\B\in\Pi^\prime(G,H)}\tp{\vect(\C_k)}\vect(\B)$.\label{ipfp-2}
\item Set $\X^\prime\coloneqq\argmin_{\X\in\{\X^\prime,\B_{k+1}\}}c(\X)$.\label{ipfp-3}
\item Compute $\alpha_{k+1} \coloneqq \min_{\alpha\in[0,1]} c(\X_k+ \alpha\cdot(\B_{k+1}-\X_k))$.\label{ipfp-4}
\item Set $\X_{k+1}\coloneqq\X_k + \alpha_{k+1} (\B_{k+1} - \X_k)$.\label{ipfp-5}
\end{enumerate}

\IPFP iterates until $|c(\X_k)-\tp{\vect(\C_k)}\vect(\B_{k+1})|/c(\X_k)$ is smaller than a threshold $\varepsilon$ or a maximal number of iterations $I$ has been reached. Subsequently, the local optimum $\X_{k+1}$ is projected to the closest integral solution $\widehat{\X}$, and the best encountered node map $\X^\prime\coloneqq\argmin_{\X\in\{\X^\prime,\widehat{\X}\}}c(\X)$ is returned.


\subsubsection{The algorithm \BPBEAM}\label{sec:related:bp-beam}

Given an initial node map $\pi\in\Pi(G,H)$ and a constant $B\in\N_{\geq1}$, the algorithm \BPBEAM \cite{riesen:2014aa} starts by producing a random ordering $((u_s,v_s))^{|\pi|}_{s=1}$ of the initial node map $\pi$. \BPBEAM now constructs an improved node map $\pi^\prime$ by partially traversing an implicitly constructed tree $T$ via beam search with beam size $B$. The nodes of $T$ are tuples $(\pi^{\prime\prime},s)$, where $\pi^{\prime\prime}\in\Pi(G,H)$ is an ordered node map and $s\in[|\pi|]$ is the depth of the tree node in $T$. Tree nodes $(\pi^{\prime\prime},s)$ with $s=|\pi|$ are leafs, and the children of an inner node $(\pi^{\prime\prime},s)$ are $\{(\SWAP(\pi^{\prime\prime},s,s^\prime),s+1)\mid s^\prime\in\{s,\ldots,|\pi|\}\}$. Here, $\SWAP(\pi^{\prime\prime},s,s^\prime)$ is the ordered node map obtained from $\pi^{\prime\prime}$ by swapping the assignments $(u_s,v_s)$ and $(u_{s^\prime},v_{s^\prime})$.

At initialization, \BPBEAM sets the output node map $\pi^\prime$ to the initial node map $\pi$. Furthermore, \BPBEAM maintains a priority queue $q$ of tree nodes which is initialized as $q\coloneqq\{(\pi,1)\}$ and sorted \wrt non-decreasing induced edit cost of the contained node maps. As long as $q$ is non-empty, \BPBEAM extracts the top node $(\pi^{\prime\prime},s)$ from $q$ and updates the output node map $\pi^\prime$ to 
$\pi^{\prime\prime}$ if $c(P_{\pi^{\prime\prime}})<c(P_{\pi^\prime})$. If $s<|\pi|$, \ie, if the extracted tree node is no leaf, \BPBEAM adds all of its children to $q$ and subsequently discards all but the first $B$ tree nodes contained in $q$. Once $q$ is empty, the cheapest encountered node map $\pi^\prime$ is returned.

\subsection{The algorithm \IBPBEAM}\label{sec:related:ibp-beam}

Since the size of the priority queue $q$ is restricted to $B$, which parts of the search tree $T$ are visited by \BPBEAM crucially depends on the ordering of the initial node map $\pi$. Therefore, \BPBEAM can be improved by considering not one but several initial orderings. The algorithm \IBPBEAM suggested in \cite{ferrer:2015ab} does exactly this. That is, given a constant number of iterations $I\in\N_{\geq1}$, \IBPBEAM runs \BPBEAM with $I$ different randomly created orderings of the initial node map $\pi$, and then returns the cheapest node map $\pi^\prime$ encountered in one of the iterations.

\section{The local search algorithm \KREFINE}\label{sec:k-refine}

In this section, we extend and improve the algorithm \REFINE \cite{zeng:2009aa} in three ways. Firstly, instead of considering only binary swaps, we make \KREFINE consider all $K^\prime$-swaps for all $K^\prime\in[K]\setminus\{1\}$, where $K\in\N_{\geq2}$ is a constant (\Cref{sec:k-refine:k-swap}). Secondly, we show that for computing the induced cost $c(P_{\pi^\prime})$ of a node map $\pi^\prime$ obtained from $\pi$ via a $K^\prime$-swap $C$, it suffices to consider the nodes and edges that are incident with $C$ (\Cref{sec:k-refine:swap-costs}). This observation yields an improved implementation of \KREFINE, which is much more efficient than the na\"{i}ve implementation suggested in \cite{zeng:2009aa}. Thirdly, we suggest to include the dummy assignment $(\epsilon,\epsilon)$ into the initial node map $\pi$ before enumerating the swaps (\Cref{sec:k-refine:dummy-assignment}). This modification allows the number of node substitutions to decrease and hence improves the quality of the obtained upper bound.

\subsection{Generalization to swaps of size larger than two}\label{sec:k-refine:k-swap}

\Cref{algo:k-refine} gives an overview of the algorithm \KREFINE, which generalizes \REFINE to swaps of size larger than two. Given graphs $G$ and $H$, an initial node map $\pi\in\Pi(G,H)$, and a maximal swap size $K\in\N_{\geq2}$, \KREFINE starts by initializing the current swap size $K^\prime$, the best swap $C^\star$, and the best swap cost $\Delta^\star$ as $K^\prime\coloneqq2$, $C^\star\coloneqq\emptyset$, and $\Delta^\star\coloneqq0$ (\crefrange{algo:k-refine:init-swap-size}{algo:k-refine:init-best-swap}). Subsequently, \KREFINE enters its main while-loop and iterates until no improved node map has been found and the current swap size exceeds the maximal swap size (\cref{algo:k-refine:main-loop}).

\begin{algorithm}[t]
\SetAlFnt{\small}
\SetCommentSty{emph}
\SetKwInput{Out}{Output}
\SetKwInput{In}{Input}
\In{Graphs $G$ and $H$, initial node map $\pi\in\Pi(G,H)$, constant $K\in\N_{\geq2}$.}
\Out{Node map $\pi^\prime\in\Pi(G,H)$ with $c(P_{\pi^\prime})\leq c(P_\pi)$.}
\BlankLine
$K^\prime\coloneqq2$\label{algo:k-refine:init-swap-size}\tcp*{initialize current size}
$C^\star\coloneqq\emptyset$; $\Delta^\star\coloneqq0$\label{algo:k-refine:init-best-swap}\tcp*{initialize best swap}
\While(\tcp*[f]{main loop}){$\Delta^\star<0\lor K^\prime\leq K$\label{algo:k-refine:main-loop}}{
\For(\tcp*[f]{enumerate swaps of current size}){$C\in\mathcal{C}_{\pi,K^\prime}$\label{algo:k-refine:enumerate-swaps}}{
$\Delta\coloneqq\SWAPCOST(\pi,C)$\label{algo:k-refine:swap-cost}\tcp*{compute swap cost}
\If(\tcp*[f]{found better swap}){$\Delta<\Delta^\star$\label{algo:k-refine:found-better-swap}}{
$C^\star\coloneqq C$; $\Delta^\star\coloneqq\Delta$\label{algo:k-refine:update-best-swap}\tcp*{update best swap}
}
}
\eIf(\tcp*[f]{found better node map}){$\Delta^\star<0$\label{algo:k-refine:found-better-pi}}{
$\pi^\prime\coloneqq\SWAP(\pi,C^\star,)$\label{algo:k-refine:update-pi-start}\tcp*{compute swapped node map}
$c(P_{\pi^\prime})\coloneqq c(P_\pi)+\Delta^\star$\tcp*{set swapped node map cost}
$\pi\coloneqq\pi^\prime$\label{algo:k-refine:update-pi-end}\tcp*{update current node map}
$K^\prime\coloneqq2$\label{algo:k-refine:reset-swap-size}\tcp*{reset current swap size}
}{
$K^\prime\coloneqq K^\prime+1$\label{algo:k-refine:increment-swap-size}\tcp*{increment current swap size}
}
$C^\star\coloneqq\emptyset$; $\Delta^\star\coloneqq0$\label{algo:k-refine:reset-best-swap}\tcp*{reset best swap}
}
\Return $\pi^\prime\coloneqq\pi$\label{algo:k-refine:return-node-map}\tcp*{return improved node map}
\caption{The algorithm \KREFINE.}\label{algo:k-refine}
\end{algorithm}
%
%

Inside the main while-loop, the algorithm \KREFINE first enumerates the set $\mathcal{C}_{\pi,K^\prime}\coloneqq\{C\subseteq A_\pi\mid\text{$C$ is cycle of length $K^\prime$ in $G_\pi$}\}$ of all $K^\prime$-swaps of $\pi$ (\cref{algo:k-refine:enumerate-swaps}). The auxiliary directed bipartite graph $G_\pi\coloneqq(V^G_\pi\cup V^H_\pi,A_\pi)$ is defined as in \Cref{sec:related:refine} above, \ie, we have $V^G_\pi\coloneqq\{u_s\mid s\in[|\pi|]\}$, $V^H_\pi\coloneqq\{v_s\mid s\in[|\pi|]\}$, and $A_\pi\coloneqq\pi\cup\{(v_s,u_{s^\prime})\in V^H_\pi\times V^G_\pi\mid (s,s^\prime)\in[|\pi|]\times[|\pi|]\land s\neq s^\prime\}$, where $((u_s,v_s))^{|\pi|}_{s=1}$ is an arbitrary ordering of $\pi$. 

For each $K^\prime$-swap $C\in\mathcal{C}_{\pi,K^\prime}$, let $F(C)\coloneqq C\cap\pi$ and $B(C)\coloneqq\{(u,v)\in(V^G\cup\{\epsilon\})\times(V^H\cup\{\epsilon\})\mid(v,u)\in C\setminus\pi\}$ be the sets of node assignments corresponding to forward and backwards arcs contained in $C$, respectively. \KREFINE computes the swap cost
\begin{IEEEeqnarray}{c}\label{eq:def-swap-cost}
\SWAPCOST(\pi,C)\coloneqq c(P_\pi)-c(P_{\SWAP(\pi,C)})\text{,}
\end{IEEEeqnarray}
where $\SWAP(\pi,C)\coloneqq(\pi\setminus F(C))\cup \{(u,v)\in B(C)\mid(u,v)\neq(\epsilon,\epsilon)\}$ is the node map obtained from $\pi$ by carrying out the swap encoded by $C$ (\cref{algo:k-refine:swap-cost}). If $C$ yields an improvement, \KREFINE updates the best swap $C^\star$ and the best swap cost $\Delta^\star$ (\crefrange{algo:k-refine:found-better-swap}{algo:k-refine:update-best-swap}). 

Once all $K^\prime$-swaps $C\in\mathcal{C}_{\pi,K^\prime}$ have been visited, \KREFINE checks whether one of them yields an improvement \wrt the current node map $\pi$ (\cref{algo:k-refine:found-better-pi}). If this is the case, \KREFINE updates $\pi$ (\crefrange{algo:k-refine:update-pi-start}{algo:k-refine:update-pi-end}) and resets the current swap size to $K^\prime\coloneqq2$ (\cref{algo:k-refine:reset-swap-size}). Otherwise, $K^\prime$ is incremented (\cref{algo:k-refine:increment-swap-size}). Subsequently, \KREFINE resets the best swap and the best swap cost to $C^\star\coloneqq\emptyset$ and $\Delta^\star\coloneqq0$, respectively (\cref{algo:k-refine:reset-best-swap}). Upon termination of the main loop, \KREFINE returns the current node map $\pi$ (\cref{algo:k-refine:return-node-map}).

Assume that $\SWAPCOST(\pi,C)$ can be computed in $O(\omega)$ time (\cf \Cref{sec:k-refine:swap-costs} for details). Furthermore, let $I\in\N$ be the number of times \KREFINE finds an improved node map in \cref{algo:k-refine:found-better-pi}. Note that, if the edit costs are integral, it holds that $I\leq c(P_\pi)$, where $\pi$ is \KREFINE's initial node map. \Cref{prop:num-swaps} below implies that, for all $K^\prime\in[K]\setminus\{1\}$ and each node map $\pi\in\Pi(G,H)$, we have $|\mathcal{C}_{\pi,K^\prime}|=O((|V^G|+|V^H|)^{K^\prime})$. Therefore, \KREFINE's overall runtime complexity is $O(I(|V^G|+|V^H|)^K\omega)$.

\begin{proposition}\label{prop:num-swaps}
For each node map $\pi\in\Pi(G,H)$ and each $K^\prime\in\N_{\geq2}$, it holds that $|\mathcal{C}_{\pi,K^\prime}|=\binom{|\pi|}{K^\prime}(K^\prime-1)!$.
\end{proposition}

\begin{proof}
The proposition immediately follows from the definition of $\mathcal{C}_{\pi,K^\prime}$. Details are omitted due to space constraints.
\end{proof}

\subsection{Efficient computation of swap costs}\label{sec:k-refine:swap-costs}

Given a node map $\pi\in\Pi(G,H)$ and a $K^\prime$-swap $C\in\mathcal{C}_{\pi,K^\prime}$, let $\pi^\prime\coloneqq\SWAP(\pi,C)$ be the node map obtained from $\pi$ by swapping the forward and backward arcs contained in $C$. Assume that $c(P_\pi)$ has already been computed. By \cref{eq:def-swap-cost}, the swap cost $\SWAPCOST(\pi,C)$ can be computed na\"{i}vely by computing the induced costs $c(P_{\pi^\prime})$ of the swapped node map and then considering the difference between $c(P_\pi)$ and $c(P_{\pi^\prime})$. By definition of $c(P_\pi)$, this requires $O(\max\{|E^G|,|E^H|\})$ time. Since $\SWAPCOST(\pi,C)$ has to be computed in every iteration of \KREFINE's inner for-loop, it is highly desirable to implement $\SWAPCOST(\cdot,\cdot)$ more efficiently. The following \Cref{prop:swaps-costs} provides the key ingredient of a more efficient implementation.

\begin{proposition}\label{prop:swaps-costs}
Let $\pi\in\Pi(G,H)$ be a node map, $K^\prime\in\N_{\geq2}$ be a constant, and $C\in\mathcal{C}_{\pi,K^\prime}$ be a $K^\prime$-swap. Furthermore, let $V^G_C\coloneqq\{u\in V^G\mid\exists v\in V^H\cup\{\epsilon\}:(u,v)\in F(C)\}$, $V^H_C\coloneqq\{v\in V^H\mid\exists u\in V^G\cup\{\epsilon\}:(u,v)\in F(C)\}$, $E^G_C\coloneqq\{e\in E^G\mid e\cap V^G_C\neq\emptyset\}$, and $E^H_C\coloneqq\{f\in E^H\mid f\cap V^H_C\neq\emptyset\}$, $\Delta\coloneqq\SWAPCOST(\pi,C)$, and $\pi^\prime\coloneqq\SWAP(\pi,C)$. Then the following equation holds:
\begin{IEEEeqnarray*}{rCl}
\Delta&=&\sum_{\substack{u\in V^G_C\\\pi^\prime(u)\neq\epsilon}}c_V(u,\pi^\prime(u))+\sum_{\substack{u\in V^G_C\\\pi^\prime(u)=\epsilon}}c_V(u,\epsilon)+\sum_{\substack{v\in V^H_C\\{\pi^\prime}^{-1}(v)=\epsilon}}c_V(\epsilon,v)\\
&&+\>\sum_{\substack{e\in E^G_C\\\pi^\prime(e)\neq\epsilon}}c_E(e,\pi^\prime(e))+\sum_{\substack{e\in E^G_C\\\pi^\prime(e)=\epsilon}}c_E(e,\epsilon)+\sum_{\substack{f\in E^H_C\\{\pi^\prime}^{-1}(f)=\epsilon}}c_E(\epsilon,f)\\
&&-\>\sum_{\substack{u\in V^G_C\\\pi(u)\neq\epsilon}}c_V(u,\pi(u))-\sum_{\substack{u\in V^G_C\\\pi(u)\neq\epsilon}}c_V(u,\pi(u))-\sum_{\substack{v\in V^H_C\\\pi^{-1}(v)=\epsilon}}c_E(\epsilon,v)\\
&&-\>\sum_{\substack{e\in E^G_C\\\pi(e)\neq\epsilon}}c_E(e,\pi(e))-\sum_{\substack{e\in E^G_C\\\pi(e)=\epsilon}}c_E(e,\epsilon)-\sum_{\substack{f\in E^H_C\\\pi^{-1}(f)=\epsilon}}c_E(\epsilon,f)
\end{IEEEeqnarray*}
\end{proposition}

\begin{proof}
By construction of $V^G_C$ and $V^H_C$, we have $\pi(u)=\pi^\prime(u)$, for all $u\in V^G\setminus V^G_C$, and $\pi^{-1}(v)={\pi^\prime}^{-1}(v)$, for all $v\in V^H\setminus V^H_C$. Similarly, $\pi(e)=\pi^\prime(e)$ and $\pi^{-1}(f)={\pi^\prime}^{-1}(f)$ holds for all $e\in E^G\setminus E^G_C$ and all $f\in E^H\setminus E^H_C$. This proves the proposition.
\end{proof}

\Cref{prop:swaps-costs} implies that, for computing $\SWAPCOST(\pi,C)$, only the nodes and edges contained in $V^G_C$, $V^H_C$, $E^G_C$, and $E^H_C$ must be considered. Since $K^\prime$ is constant, $|V^G_C|,|V^H_C|\leq K^\prime$, $|E^G_C|\leq K^\prime\maxdeg(G)$, and $|E^H_C|\leq K^\prime\maxdeg(H)$, $\SWAPCOST(\pi,C)$ can hence be computed in $O(\max\{\maxdeg(H),\maxdeg(G)\})$ time. This is a significant improvement \wrt the na\"{i}vely computing $\SWAPCOST(\pi,C)$ in $O(\max\{|E^G|,|E^H|\})$ time.

\subsection{Improvement of upper bound via dummy assignment}\label{sec:k-refine:dummy-assignment}

For each node map $\pi\in\Pi(G,H)$, let $S(\pi)\coloneqq|\{(u,v)\in\pi\mid u\neq\epsilon\land v\neq\epsilon\}|$ denote the number of node substitutions contained in $\pi$. Now assume that \KREFINE as specified in \Cref{algo:k-refine} is run from an initial node map $\pi\in\Pi(G,H)$ that does not contain the dummy assignment $(\epsilon,\epsilon)$. Since $\pi$ and $\pi\cup\{(\epsilon,\epsilon)\}$ induce the same edit path, this assumption is likely to hold in most implementations of \KREFINE. The following \Cref{prop:k-refine:dummy-assignment} shows that, under this assumption, the search space of \KREFINE is restricted in the sense that it includes only node maps $\pi^\prime\in\Pi(G,H)$ with $S(\pi^\prime)\geq S(\pi)$. This has a negative effect on the quality of the upper bound produced by \KREFINE, as some potentially promising node maps are excluded a priori.

\begin{proposition}\label{prop:k-refine:dummy-assignment}
Let $\pi\in\Pi(G,H)$ be a node map that satisfies $(\epsilon,\epsilon)\notin\pi$ and $\pi^\prime\in\Pi(G,H)$ be the improved node map obtained from $\pi$ by running \KREFINE as specified in \Cref{algo:k-refine}. Then it holds that $S(\pi^\prime)\geq S(\pi)$.
\end{proposition}

\begin{proof}
Let $\pi\in\Pi(G,H)$ be a node map that satisfies $(\epsilon,\epsilon)\notin\pi$, $K^\prime\in\N_{\geq2}$ be a constant, and $C\in\mathcal{C}_{\pi,K^\prime}$ be a $K^\prime$-swap. By definition of $B(C)$, we have $(\epsilon,\epsilon)\notin\SWAP(\pi,C)$. Therefore, the proposition follows by induction on the number of times \KREFINE finds an improved node map in \cref{algo:k-refine:found-better-pi}, if we can show that $S(\SWAP(\pi,C))\geq S(\pi)$. To show this inequality, we define $S^\epsilon_F(C)\coloneqq|\{(u,v)\in F(C)\mid u=\epsilon\land v=\epsilon\}|$ and $S^\epsilon_B(C)\coloneqq|\{(u,v)\in B(C)\mid u=\epsilon\land v=\epsilon\}|$. It is easy to see that we have $S(\SWAP(\pi,C))=S(\pi)+S^\epsilon_B(C)-S^\epsilon_F(C)$. Since $(\epsilon,\epsilon)\notin\pi$, we additionally know that $S^\epsilon_F(C)=0$. We hence obtain $S(\SWAP(\pi,C))=S(\pi)+S^\epsilon_B(C)\geq S(\pi)$, as required.
\end{proof}

The proof of \Cref{prop:k-refine:dummy-assignment} tells us how we have to modify \KREFINE in order to ensure that node maps with fewer node substitutions than the initial node map are contained its search space: We simply have to update the current node map $\pi$ as $\pi\coloneqq\pi\cup\{(\epsilon,\epsilon)\}$ before enumerating all $K^\prime$-swaps $C\in\mathcal{C}_{\pi,K^\prime}$ in \cref{algo:k-refine:enumerate-swaps} of \Cref{algo:k-refine}. 
This modification is particularly important if the edit costs are non-metric, \ie, if it can happen that deleting plus inserting is cheaper than substituting.

\section{The framework \RANDPOST}\label{sec:randpost}

In this section, we present \RANDPOST, a framework that can be used to improve any local search based algorithm for upper bounding \GED. Intuitively, \RANDPOST iteratively runs a given local search algorithm. In each iteration, previously computed locally optimal node maps are combined stochastically to obtain new promising initial node maps to be used in the next iteration.

\Cref{algo:randpost} provides an overview of the framework. Given a set of initial node maps $\mathcal{S}_0\subseteq\Pi(G,H)$ with $|\mathcal{S}_0|=\kappa$, a constant $\rho\in(0,1]$, and a local search algorithm \ALG, \RANDPOST computes a set $\mathcal{S}^\prime_0\subseteq\Pi(G,H)$ of improved node maps with $|\mathcal{S}^\prime_0|=\ceil*{\rho\cdot \kappa}$ by (parallelly) running \ALG on all initial node maps and terminating once $\ceil*{\rho\cdot \kappa}$ runs have converged (\cref{algo:randpost:run-init-ls}). Subsequently, the upper bound \UB is initialized as the cost of the cheapest induced edit path encountered so far (\cref{algo:randpost:set-first-ub}). Note that, up to this point, \RANDPOST is equivalent to the local search framework with multi-start described in \Cref{sec:prelim:paradigm} above.

\begin{algorithm}[t]
\SetAlFnt{\small}
\SetCommentSty{emph}
\SetKwInput{Out}{Output}
\SetKwInput{In}{Input}
\In{Graphs $G$ and $H$, constants $\kappa\in\N_{\geq1}$, $L\in\N$, $\rho\in(0,1]$, and $\eta\in[0,1]$, local search algorithm \ALG, initial node map set $\mathcal{S}_0\subseteq\Pi(G,H)$ with $|\mathcal{S}_0|=\kappa$, lower bound $\LB$ for $\GED(G,H)$.}
\Out{Upper bound \UB for $\GED(G,H)$.}
\BlankLine
$\mathcal{S}^\prime_0\coloneqq\ALG(\mathcal{S}_0,\rho)$\label{algo:randpost:run-init-ls}\tcp*{run local search on initial node maps}
$\UB\coloneqq\min_{\pi^\prime \in\mathcal{S}^\prime_0}c(P_{\pi^\prime})$\label{algo:randpost:set-first-ub}\tcp*{set first upper bound}
$\M\coloneqq\mathbf{0}_{(|V^G|+1)\times(|V^H|+1)}$\label{algo:randpost:init-scores}\tcp*{initialize scores matrix}
\For(\tcp*[f]{main loop}){$r\in[L]$\label{algo:randpost:main-loop}}{
$\M\coloneqq\UPDATESCORES(\M,\mathcal{S}^\prime_{r-1},\eta,\LB,\UB)$\label{alg:randpost:update-score}\tcp*{update scores}
$\mathcal{S}_r\coloneqq\GENERATENODEMAPS(\M,\kappa)$\label{algo:randpost:generate-node-maps}\tcp*{generate node maps}
$\mathcal{S}^\prime_r\coloneqq\ALG(\mathcal{S}_r,\rho)$\label{algo:randpost:run-next-ls}\tcp*{run local search on new node maps}
$\UB\coloneqq\min\{\UB,\min_{\pi^\prime\in\mathcal{S}^\prime_r}c(P_{\pi^\prime})\}$\label{algo:randpost:update-ub}\tcp*{update upper bound}  
}
\Return \UB\label{algo:randpost:return-ub}\tcp*{return upper bound} 
\caption{The framework \RANDPOST.}\label{algo:randpost}
\end{algorithm}

\RANDPOST now initializes a matrix $\M\in\R^{(|V^G|+1)\times(|V^H|+1)}\coloneqq\mathbf{0}_{(|V^G|+1)\times(|V^H|+1)}$ that contains scores $m_{i,k}$ for each possible node assignment $(u_i,v_k)\in(V^G\cup\{\epsilon\})\times(V^H\cup\{\epsilon\})$ (\cref{algo:randpost:init-scores}). The score for each substitution $(u_i,v_k)\in V^G\times V^H$ is represented by the value $m_{i,k}$, while the scores for the deletion $(u_i,\epsilon)$ and the insertion $(\epsilon,v_k)$ are represented by the values $m_{i,|V^H|+1}$ and $m_{|V^G|+1,k}$, respectively. Throughout the algorithm, \M is maintained in such a way that $m_{i,k}$ is large just in case the corresponding node assignment appears in many cheap locally optimal node maps.

After initializing \M, \RANDPOST carries out $L$ iterations of its main for-loop, where $L\in\N$ is a meta-parameter (\crefrange{algo:randpost:main-loop}{algo:randpost:update-ub}). Inside the $r$\textsuperscript{th} iteration, \RANDPOST starts by updating the scores matrix \M by calling $\UPDATESCORES(\M,\mathcal{S}^\prime_{r-1},\eta,\LB,\UB)$, where \M is the current scores matrix, $\mathcal{S}^\prime_{r-1}\subseteq\Pi(G,H)$ is the set of improved node maps obtained from the previous iteration, $\eta\in[0,1]$ is a meta-parameter used to give greater weight to cheap node maps, \LB is a previously computed lower bound for $\GED(G,H)$, and \UB is the current upper bound (\cref{alg:randpost:update-score}). Let the matrix $\M^\prime\in\R^{(|V^G|+1)\times(|V^H|+1)}$ be defined as $\M^\prime\coloneqq\UPDATESCORES(\M,\mathcal{S}^\prime_{r-1},\eta,\LB,\UB)$. Then $\M^\prime$ is given as
\begin{IEEEeqnarray*}{c}
\M^\prime\coloneqq\M+\sum_{\pi^\prime\in\mathcal{S}^\prime_{r-1}}\left[(1-\eta)+\eta\frac{\UB-\LB}{c(P_{\pi^\prime})-\LB}\right]\X^\prime\text{,}
\end{IEEEeqnarray*} 
where $\X^\prime\in\{0,1\}^{(|V^G|+1)\times(|V^H|+1)}$ is the matrix representation of the improved node map $\pi^\prime\in\mathcal{S}^\prime_{r-1}$, \ie, for all $u_i\in V^G$ and all $v_k\in V^H$, we have $x^\prime_{i,k}=1$ just in case $(u_i,v_k)\in\pi^\prime$, $x^\prime_{i,|V^H|+1}=1$ just in case $(u_i,\epsilon)\in\pi^\prime$, and $x^\prime_{|V^G|+1,k}=1$ just in case $(\epsilon,v_k)\in\pi^\prime$. If $\eta=0$, $m_{i,k}$ represents the number of converged local optima that contain the corresponding assignment. If $\eta>0$, assignments that appear in node maps with lower costs receive higher scores.

Once \M has been updated, \RANDPOST creates a new $\kappa$-sized set $\mathcal{S}_r\subseteq\Pi(G,H)$ of initial node maps by calling $\GENERATENODEMAPS(\M,\kappa)$ (\cref{algo:randpost:generate-node-maps}). $\GENERATENODEMAPS(\M,\kappa)$ works as follows: For each of the first $|V^G|$ rows $\M_i$ of \M, \RANDPOST draws a column $k\in[|V^H|+1]$ from the distribution encoded my $\M_i$. If $k=|V^H|+1$, the node deletion $(u_i,\epsilon)$ is added to the node map $\pi$ that is being constructed. Otherwise, the substitution $(u_i,v_k)$ is added to $\pi$, the score $m_{j,k}$ is temporarily set to $0$ for all $j\in[|V^G|]\setminus[i]$, and the column $k$ is marked as covered. Once all nodes of $G$ have been processed, node insertions $(\epsilon,v_k)$ are added to $\pi$ for all uncovered columns $k\in[|V^H|]$. This process is repeated until $\kappa$ different node maps have been created.

After creating the set $\mathcal{S}_r$ of new initial node maps, \RANDPOST constructs a new $\ceil*{\rho\cdot \kappa}$-sized set $\mathcal{S}^\prime_r\subseteq\Pi(G,H)$ of improved node maps by (parallelly) running the local search algorithm \ALG on the initial node maps contained in $\mathcal{S}_r$ and terminating once $\ceil*{\rho\cdot \kappa}$ runs have converged (\cref{algo:randpost:run-next-ls}). Subsequently, the upper bound is updated as the minimum of the current upper bound and the cost of the cheapest edit path induced by one of the newly computed improved node maps (\cref{algo:randpost:update-ub}). Finally, \RANDPOST returns the best encountered upper bound (\cref{algo:randpost:return-ub}).

\begin{table}[!t]
\centering
\caption{Properties of test datasets.}\label{tab:datasets}
\begin{small}
\begin{tabular}{lcccc}
\toprule
& \grec & \fingerprint & \protein & \mutagenicity \\
\midrule
max./avg.\@ $|V^G|$ & 11.5/26 & 5.4/26 & 126/32.6 & 30.3/417 \\
max./avg.\@ $|E^G|$ & 12.2/30 & 4.4/25 & 149/62.1 & 30.8/112 \\
\bottomrule
\end{tabular}
\end{small}
\end{table}
\section{Empirical evaluation}\label{sec:experiments}

In order to empirically evaluate \KREFINE and \RANDPOST, extensive tests were conducted on four standard datasets from the IAM Database Repository \cite{riesen:2008aa,riesen:2010aa}: \mutagenicity, \protein, \grec, and \fingerprint (\cf \Cref{tab:datasets}). For all datasets, we tested on the metric edit costs suggested in \cite{riesen:2010aa}. For \mutagenicity, we additionally defined non-metric edit costs by setting the costs of node and edge deletions and insertions to $1$, and setting the costs of node and edge substitutions to $3$ (the resulting dataset is denoted as \mutagenicityn). For each dataset, subsets of 50 graphs were chosen randomly, and upper bounds for \GED were computed for each pair of graphs in the subsets, as well as for each graph and a shuffled copy of itself. In the following, $d$, $\hat{d}$, and $t$ denote the average upper bound, the average upper bound between graphs and their shuffled copies, and the average runtime in seconds, respectively. Note that the test metric $\hat{d}$ gives us a hint to how close to optimality each algorithm is, as the optimal value, namely 0, is known. All methods were implemented using the GEDLIB library and were run in 20 parallel threads.\footnote{Sources and datasets: \url{https://github.com/dbblumenthal/gedlib/}.}

We tested two versions \KREFINE[2] and \KREFINE[3] of our local search algorithm \KREFINE, which use swaps of maximum size two and three, respectively. We compared them to the existing local search algorithms \REFINE, \IPFP, \BPBEAM, and \IBPBEAM. As suggested in \cite{riesen:2014aa} and \cite{ferrer:2015ab}, we set the beam size employed by \BPBEAM and \IBPBEAM to 5 and the number of iterations employed by \IBPBEAM to 20. \IPFP was run with convergence threshold set to \num{E-3} and maximum number of iterations set to 100, as proposed in \cite{blumenthal:2018ac}. In order to evaluate \RANDPOST, we ran each algorithm with $\kappa\coloneqq40$ initial solutions. Indeed, experiments reported in~\cite{daller:2018aa} show that on all tested datasets, using more initial solutions does not bring a significant decrease of the estimated GED. We also varied the pair of meta-parameters $(L,\rho)$ on the set $\{(0,1),(1,0.5),(3,0.25),(7,0.125)\}$. Recall that $L$ is the number of \RANDPOST loops and $\rho$ is defined such that each iteration produces exactly $\ceil*{\rho\cdot \kappa}$ locally optimal node maps. Therefore, our setup ensures that each configuration produces exactly 40 local optima. For each algorithm and each dataset, we conducted pre-tests where we varied the penalty parameter $\eta$ on the set $\{n/10\mid n\in\N_{\leq10}\}$, and then picked the value of $\eta$ for the main experiments that yielded the best average upper bound across all \RANDPOST configurations.

\subsection{\KREFINE vs.\@ \REFINE}\label{sec:experiments:k-refine}

In a first series of experiments, we compared the versions \KREFINE[2] and \KREFINE[3] of our improved and generalized local search algorithm \KREFINE to the baseline algorithm \REFINE. All algorithms were run without \RANDPOST, \ie, with $(L,\rho)=(0,1)$, and the tests were carried out  on a computer using an Intel Xeon E5-2620 v4 2.10GHz CPU. \Cref{tab:k-refine} shows the results. By comparing $t(\KREFINE[2])$ and $t(\REFINE)$, we see that efficiently computing the swap costs as suggested in \Cref{sec:k-refine:swap-costs} indeed significantly improves the runtime performance. Unsurprisingly, the speed-up is especially large on the datasets \protein and \mutagenicity containing the larger graphs. Comparing $d(\KREFINE[2])$ and $d(\REFINE)$ shows that the inclusion of the dummy assignment proposed in \Cref{sec:k-refine:dummy-assignment} slightly improves the quality of the produced upper bound. As expected, the percentual improvement is largest on the dataset \mutagenicityn with non-metric edit costs. Finally, we observe that running \KREFINE with swaps of size three slightly improves the upper bounds on all datasets, but significantly increases the runtime of the algorithm. 

\begin{table}[!t]
\caption{\KREFINE vs.\@ \REFINE without \RANDPOST.}\label{tab:k-refine}
\centering
\begin{small}
\begin{tabular}{lS[table-format=3.2]S[table-format=1.2e-1]S[table-format=3.2]S[table-format=1.2e-1]S[table-format=3.2]S[table-format=1.2e-1]}
\toprule
dataset & \multicolumn{2}{c}{\REFINE} & \multicolumn{2}{c}{\KREFINE[2]} & \multicolumn{2}{c}{\KREFINE[3]} \\
\midrule
& {$d$} & {$t$} & {$d$} & {$t$} & {$d$} & {$t$} \\
\cmidrule(rl){2-3} \cmidrule(rl){4-5} \cmidrule(l){6-7}
\grec & 859.92 & 2.46e-2 & 857.89 & \bfseries 1.15e-2 & \bfseries 857.12 & 3.85e-2 \\ 
\fingerprint & \bfseries 2.82 & 5.34e-4 & \bfseries 2.82 & \bfseries 4.49e-4 & \bfseries 2.82 & 1.58e-3 \\ 
\protein & 295.61 & 3.43e-1 & 295.55 & \bfseries 9.07e-2 & \bfseries 295.29 & 4.89e-1 \\
\mutagenicity & 74.12 & 1.22e-1 & 74.12 & \bfseries 3.92e-2 & \bfseries 73.61 & 1.87e-1 \\
\mutagenicityn & 49.49 & 1.14e-1 & 49.11 & \bfseries 3.58e-2 & \bfseries 48.44 & 1.81e-1 \\
\bottomrule
\end{tabular}
\end{small}
\end{table}

\subsection{Behaviour of \RANDPOST framework}\label{sec:experiments:randpost}
\begin{figure}[!t]
\centering
\begin{tikzpicture}
\begin{groupplot}
[
group style={group name=lineplots, group size=3 by 5, horizontal sep=1cm, vertical sep=1cm},
width=.4\linewidth,
height=.4\linewidth,
legend columns=5,
legend cell align=left,
legend style={align=left, draw=none, column sep=.5ex, font=\footnotesize}
]
\addplots{\fingerprint}{FP_pgf.csv}{avg_runtime}{$t$ in sec.}{normal}
\addplots{\fingerprint}{FP_pgf.csv}{avg_ub}{$d$}{normal}
\addplots{\fingerprint}{FP_pgf.csv}{avg_ub_diag}{$\hat{d}$}{normal}
\addplots{\grec}{GREC_pgf.csv}{avg_runtime}{$t$ in sec.}{normal}
\addplots{\grec}{GREC_pgf.csv}{avg_ub}{$d$}{normal}
\addplots{\grec}{GREC_pgf.csv}{avg_ub_diag}{$\hat{d}$}{normal}
\addplots{\protein}{PROTEIN_pgf.csv}{avg_runtime}{$t$ in sec.}{normal}
\addplots{\protein}{PROTEIN_pgf.csv}{avg_ub}{$d$}{normal}
\addplots{\protein}{PROTEIN_pgf.csv}{avg_ub_diag}{$\hat{d}$}{normal}
\addplots{\mutagenicity}{MUTA_pgf.csv}{avg_runtime}{$t$ in sec.}{normal}
\addplots{\mutagenicity}{MUTA_pgf.csv}{avg_ub}{$d$}{normal}
\addplots{\mutagenicity}{MUTA_pgf.csv}{avg_ub_diag}{$\hat{d}$}{normal}
\addplots{\mutagenicityl}{MUTA_70_pgf.csv}{avg_runtime}{$t$ in sec.}{normal}
\addplots{\mutagenicityl}{MUTA_70_pgf.csv}{avg_ub}{$d$}{normal}
\addplots{\mutagenicityl}{MUTA_70_pgf.csv}{avg_ub_diag}{$\hat{d}$}{normal}
\end{groupplot}
\node at ($(lineplots c1r1.north) !.5! (lineplots c3r1.north)$) [inner sep=0pt,anchor=south, yshift=3ex] {\pgfplotslegendfromname{grouplegend}};
\end{tikzpicture}
\caption{Effect of \RANDPOST on local search algorithms.}\label{fig:randpost}
\end{figure}
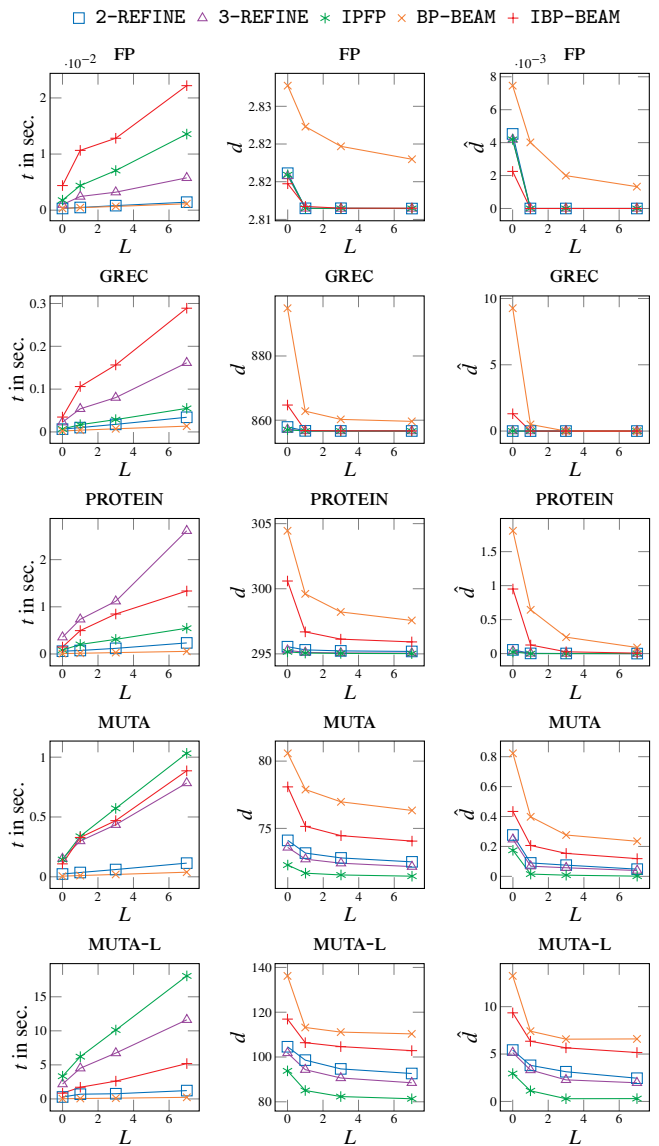

\begin{table*}[!t]
\caption{Detailed experimental results on \mutagenicityl.}\label{tab:MUTA70}
\centering
\begin{small}
\begin{tabular}{lS[table-format=3.2]S[table-format=1.2]S[table-format=1.2e-1]S[table-format=3.2]S[table-format=1.2]S[table-format=1.2e1]S[table-format=2.2]S[table-format=1.2]S[table-format=1.2e1]S[table-format=3.2]S[table-format=2.2]S[table-format=1.2e-1]S[table-format=3.2]S[table-format=1.2]S[table-format=1.2e-1]}
\toprule
$(L,\rho)$ &  \multicolumn{3}{c}{\KREFINE[2]} & \multicolumn{3}{c}{\KREFINE[3]} & \multicolumn{3}{c}{\IPFP} & \multicolumn{3}{c}{\BPBEAM} & \multicolumn{3}{c}{\IBPBEAM} \\
\midrule
& {$d$} & {$\hat{d}$} &{$t$} & {$d$} & {$\hat{d}$} &{$t$}& {$d$} & {$\hat{d}$} &{$t$}& {$d$} & {$\hat{d}$} &{$t$}& {$d$} & {$\hat{d}$} &{$t$} \\
\cmidrule(rl){2-4} \cmidrule(rl){5-7} \cmidrule(l){8-10} \cmidrule(l){11-13} \cmidrule(l){14-16}
$(0,1)$     & 104.57 & 5.42 & 2.85e-1 & 101.82 & 5.16 & 2.18e0 & 93.81 & 2.94 & 3.34e0 & 136.21 & 13.25 & 2.95e-2 & 116.92 & 9.35 & 8.46e-1 \\
$(1,0.5)$   &  98.61 & 3.81 & 6.96e-1 &  94.34 & 3.32 & 4.51e0 & 85.05 & 1.13 & 6.21e0 & 113.18 &  7.41 & 5.81e-2 & 106.39 & 6.35 & 1.71e0 \\
$(3,0.25)$  &  94.69 & 3.13 & 7.61e-1 &  90.64 & 2.28 & 6.74e0 & 82.34 & 0.29 & 1.01e1 & 111.12 &  6.57 & 1.11e-1 & 104.64 & 5.65 & 2.61e0 \\
$(7,0.125)$ &  92.63 & 2.45 & 1.22e0  &  88.49 & 1.97 & 1.16e1 & 81.36 & 0.29 & 1.81e1 & 110.32 &  6.59 & 2.35e-1 & 102.84 & 5.16 & 5.19e0 \\
\bottomrule 
\end{tabular}
\end{small}
\end{table*}

In a second series of experiments, we evaluated the behaviour of \RANDPOST by running each algorithm with four different pairs of meta-parameters $(L,\rho)\in\{(0,1),(1,0.5),(3,0.25),(7,0.125)\}$. We remind that the case $(L,\rho)=(0,1)$ amounts to a basic multi-start framework with no \RANDPOST loop.  An additional subset of 10 graphs having exactly 70 nodes was extracted from \mutagenicity and is denoted by  \mutagenicityl. The tests were run on a computer using an Intel(R) Xeon E5-2640 v4 2.4GHz CPU. \Cref{fig:randpost} visualizes the results. Since \KREFINE[2] was always faster and more accurate than the baseline \REFINE (\cf \Cref{sec:experiments:k-refine}), we do not show plots for \REFINE. \Cref{tab:MUTA70} provides detailed numerical data for the \mutagenicityl subset, which turned out to be the subset with the highest variability in distances and computing times.

\Cref{fig:randpost} indicates that, on the datasets \fingerprint, \grec, and \protein containing small graphs,  near-optimility is reached by most algorithms when run with $L\geq1$ number of \RANDPOST loops. In these contexts, our algorithm \KREFINE[2] with \RANDPOST configuration $(L,\rho)=(1,0.5)$ provides the best tradeoff between runtime and accuracy, as it reaches the same accuracy as best algorithms, and, in terms of runtime, outperforms all algorithms except for \BPBEAM by approximately one order of magnitude. The only faster algorithm \BPBEAM computes much more expensive node maps, even in the \RANDPOST settings with higher number of loops. We also note that our algorithms \KREFINE[2] and \KREFINE[3] are already among the best local search algorithms when run in a simple multi-start setting without \RANDPOST (\ie, when $L=0$), both in terms of distance and computing time.

On the datasets \mutagenicity and \mutagenicityl containing larger graphs, the behavior of the \RANDPOST framework appears clearly and independently of the local search algorithm it is applied to. In all cases, a higher number of \RANDPOST loops\,---\,and lower number of computed solutions per loop\,---\,leads to a higher computation time (the computation time is approximately doubled whenever the number of loops is doubled), and to a lower average distance. In other words, the framework \RANDPOST provides a very useful algorithmic tool in situations where some time can be dedicated to compute tight upper bounds on big graphs.

\section{Conclusions}\label{sec:conclusions}

In this paper, we proposed \KREFINE, an improved and generalized version of the local search based \GED algorithm \REFINE, and suggested the general framework \RANDPOST, which stochastically generates promising initial solutions to tighten the upper bounds produced by all local search algorithms. Both \KREFINE and \RANDPOST perform excellently in practice: On small graphs, \KREFINE is among the algorithms computing the tightest upper bounds and, in terms of runtime, clearly outperforms all existing algorithms that yield similar accuracy. On larger graphs, \KREFINE provides a very good tradeoff between runtime and accuracy, as it is only slightly less accurate but much faster than the most accurate algorithms. The framework \RANDPOST is particularly effective on larger graphs, where it significantly improves the upper bounds of all local search algorithms.


\bibliographystyle{elsarticle-num}
\bibliography{references}

\end{document}